\documentclass{article}

\usepackage{microtype}
\usepackage{graphicx}
\usepackage[table]{xcolor}
\usepackage{subfigure}
\usepackage{booktabs} 

\usepackage{hyperref}

\definecolor{verylightgray}{gray}{0.9}

\usepackage[accepted]{icml2024}

\usepackage{amsmath}
\usepackage{amssymb}
\usepackage{mathtools}
\usepackage{amsthm}
\usepackage{multirow,tabularx,multicol}

\usepackage[capitalize,noabbrev]{cleveref}

\usepackage[textsize=tiny]{todonotes}

\newif\ifdraft
\drafttrue 
\draftfalse

\ifdraft

    \newcommand{\bsc}[1]{\textcolor{blue}{[BS: #1]}}

\else
    \newcommand{\bsc}[1]{}

\fi

\icmltitlerunning{MusicFlow: Cascaded Flow Matching for Text Guided Music Generation}

\begin{document}

\twocolumn[
\icmltitle{MusicFlow: Cascaded Flow Matching for Text Guided Music Generation}



\icmlsetsymbol{equal}{*}

\begin{icmlauthorlist}
\icmlauthor{K R Prajwal}{equal,ox}
\icmlauthor{Bowen Shi}{equal,comp}
\icmlauthor{Matthew Lee}{comp}
\icmlauthor{Apoorv Vyas}{comp}
\icmlauthor{Andros Tjandra}{comp}
\icmlauthor{Mahi Luthra}{comp}
\icmlauthor{Baishan Guo}{comp}
\icmlauthor{Huiyu Wang}{comp}
\icmlauthor{Triantafyllos Afouras}{comp}
\icmlauthor{David Kant}{comp}
\icmlauthor{Wei-Ning Hsu}{comp}
\end{icmlauthorlist}

\icmlaffiliation{ox}{VGG, University of Oxford, UK. Work done while at Meta.}
\icmlaffiliation{comp}{Meta, USA}

\icmlcorrespondingauthor{K R Prajwal}{prajwal@robots.ox.ac.uk}
\icmlcorrespondingauthor{Bowen Shi}{bshi@meta.com}

\icmlkeywords{Machine Learning, ICML}

\vskip 0.3in
]



\printAffiliationsAndNotice{\icmlEqualContribution} 

\begin{abstract}

We introduce MusicFlow, a cascaded text-to-music generation model based on flow matching. Based on self-supervised representations to bridge between text descriptions and music audios, we construct two flow matching networks to model the conditional distribution of semantic and acoustic features. Additionally, we leverage masked prediction as the training objective, enabling the model to generalize to other tasks such as music infilling and continuation in a zero-shot manner. Experiments on MusicCaps reveal that the music generated by MusicFlow exhibits superior quality and text coherence despite being over $2\sim5$ times smaller and requiring $5$ times fewer iterative steps. Simultaneously, the model can perform other music generation tasks and achieves competitive performance in music infilling and continuation. 
Our code and model will be publicly available. 
\end{abstract}

\section{Introduction}
\label{sec:intro}

Audio generation has recently received a lot of attention from the research community as well as the general public. Making sound automatically has a lot of practical applications, including voice acting, podcast making, creating foley sound effects~\cite{Luo2023DiffFoleySV}, making background music for movies~\cite{liu2023wavjourney}, and can greatly reduce the barrier for audio content creation.  In terms of research, audio generation poses a few challenges due to its long-term structure and complex interaction between channels (e.g., multiple events may appear at the same time), thus being a suitable testbed for generative models.

Modeling approaches for audio generation has rapidly progressed over the past few years due to the development of sophisticated generative methods such as autoregressive language models~\cite{Kreuk2022AudioGenTG,Wang2023NeuralCL} and non-autoregressive approaches~\cite{Le2023VoiceboxTM,Vyas2023AudioboxUA,liu2023audioldm}.
A significant portion of the generative models are focused on speech and general sound, where state-of-the-art (SOTA) models~\cite{Vyas2023AudioboxUA} are able to generate speech in diverse styles or general sound events in highly realistic manner. Compared to these two common modalities, music generation is a particularly challenging problem as it requires modeling long-term temporal structures~\cite{musiclm} and full frequency spectrum~\cite{muller2015fundamentals}. Compared to typical sound events (e.g., dog barking), it contains harmonies and melodies from different instruments. Music pieces often consist of multiple tracks, which can be intricately woven together and may involve significant interference.

With the improvement of audio tokenizers~\cite{zeghidour2021soundstream,defossez2022highfi} and generative models, the quality of generated music has been greatly improved in recent works~\cite{musiclm,musicgen}. However, many prior works are built upon language models~\cite{musiclm,musicgen,yang2023uniaudio}, which requires a computationally expensive auto-regressive inference procedure with number of forward passes proportional to the sequence length. This is worsened because many such models are based on a hierarchical set of units (e.g., Encodec tokens~\cite{musicgen}), which brings another factor up to the computation. Despite the usage of non-autoregressive models such as diffusion models~\cite{audio-ldm2,noise2music,forsgren2022riffusion,schneider2023mousai}, these approaches require hundreds of denoising steps during inference to achieve high performance. On the other hand, most of the existing models perform generation in a single stage, which models the audio waveform~\cite{noise2music} or its low-level representation such as VAE features~\cite{audio-ldm2} conditioned on text description directly. As music audios contains rich structural information and its text description can be very detailed (e.g., \emph{This is a live recording of a keyboardist playing a twelve bar blues progression on an electric keyboard. The player
adds embellishments between chord changes and the piece sounds groovy, bluesy and soulful.}), such approaches commonly fail to capture the intriguing dependency between text description and music pieces. Finally, most existing work focuses on text-to-music (TTM) generation, while lacking the ability to perform other practically useful generative tasks such as music infilling.

In this paper, we present MusicFlow, a cascaded text-to-music generation model based on flow matching. Our model is composed of two flow-matching networks, which transform text description into a sequence of semantic features and semantics into decodable acoustic features in non-autoregressive fashion. The flow matching objective equips the model with high efficiency in both training and inference, outperforming prior works with smaller model size and faster inference speed. Furthermore, by training with a masked prediction objective, MusicFlow is able to perform multiple music generation tasks, including TTM, music continuation and music infilling in a unified fashion.

%

%

%
%

\section{Related Work}
\label{sec:related}

Early works on music generation are mostly on constrained scenarios, such as generating audios for a specific style (e.g., Jazz~\cite{Hung2019ImprovingAJ}) or a specific instrument (e.g., piano~\cite{Hawthorne2018EnablingFP}). More recent works shift the focus to generating music from free-form natural language descriptions. Typically, the language description is encoded by a pre-trained text encoder, which is then used for conditioning the model. One big class of the generation backbone falls into the category of language models~\cite{musiclm,musicgen}. In this type of model, an audio is quantized into discrete units through an auto-encoder (e.g., SoundStorm~\cite{zeghidour2021soundstream}, Encodec~\cite{defossez2022highfi}). The language model is built to model the distribution of these units. During inference, the units sampled from the language model is decoded back into raw waveforms with the decoder directly without an explicit vocoder. The units are sampled either autoregressively~\cite{musicgen,musiclm,yang2023uniaudio} or in conjunction with non-autoregressive unit decoding~\cite{ziv2024masked}.  Diffusion-based music generation is typically built on top of the audio spectrogram. AudioLDM2~\cite{audio-ldm2} employs a variational auto-encoder to compress the spectrogram, where a DDIM~\cite{song2020denoising} model is trained with the compressed features. During inference, the generation is first decoded with the VAE decoder and transformed to waveform with a vocoder.  Similar approaches include Riffusion~\cite{forsgren2022riffusion}, which directly fine-tunes a stable diffusion model with spectrograms; MeLoDy~\cite{melody} which proposes a LM-guided Diffusion with a focus on fast sampling speed; and Noise2Music~\cite{noise2music}, which also builds a diffusion-based vocoder; and StableAudio~\cite{evans2024fast} which takes a latent diffusion approach, again with a focus on fast inference. 

Most of the existing methods directly learns the music distribution conditioned on text , which models the low-level audio features directly. In this work, our cascaded model is bridged by semantic features, which are learned separately with a self-supervised model. A similar approach to ours is MusicLM~\cite{musiclm}, which learns two language models generating semantic and acoustic units respectively. 
However, our model relies on flow matching, which offers improved efficiency. Its non-autoregressive nature also enables the model to better leverage context and generalize to other tasks.

\section{Method}
\label{sec:method}

\subsection{Background: Flow matching}
Introduced in~\cite{Lipman2022FlowMF}, flow matching is a method addressing continuous transformation of probability densities. Specifically, it studies flow, a time-dependent diffeomorphic mapping $\phi_t: [0,1]\times\mathbb{R}^d\rightarrow\mathbb{R}^d$, defined via the ordinary differential equation (ODE):
\begin{equation}\label{eq:ode}
    \frac{d}{dt}\phi_t(x)=v_t(\phi_t(x))
\end{equation}
$v_t:[0,1]\times\mathbb{R}^d\rightarrow\mathbb{R}^d$, namely a vector field, is parameterized by a neural network $\theta$ and learned by minimizing the flow matching objective: $L_{FM}=\mathbb{E}_{t,p_t(x)}||v_t(x;\theta)-u_t(x)||^2$, where $p_t(x)$ is a probability density path and $u_t(x)$ is the corresponding vector field. As both $p_t(x)$ and $u_t(x)$ are generally unknown, \citet{Lipman2022FlowMF} proposes minimizing the following conditional flow matching objective, which is equivalent to minimizing $L_{\text{FM}}$:
\begin{equation}\label{eq:cfm_loss}
L_{\text{CFM}}=\mathbb{E}_{t,p(x|x_1),q(x_1)}||v_t(x;\theta)-u_t(x|x_1)||^2
\end{equation}
 Considering Gaussian distributions for $p_t(x|x_1)=\mathcal{N}(x|\mu_t(x_1),\sigma_t(x_1)^2I)$, the target vector field for Equation~\ref{eq:cfm_loss} can be solved in closed form: $u_t(x|x_1)=\frac{\sigma^\prime_t(x_1)}{\sigma_t(x_1)}(x-\mu_t(x_1))+\mu^\prime_t(x_1)$. Several diffusion models~\cite{Dickstein2015deep,Ho2020Denoising,song2021scorebased} can be described under the same framework with specific conditional probability paths of $\sigma_t(x_1)$ and $\mu_t(x_1)$. Specifically, \citet{Lipman2022FlowMF} considers a conditional probability path with Gaussian mean and standard deviation changing linearly in time with $\mu_t(x)=tx$ and $\sigma_t(x)=1-(1-\sigma_{min})t$, which produces an optimal transport displacement mapping between conditional distributions. Due to its efficiency in both training and inference~\cite{Lipman2022FlowMF,Le2023VoiceboxTM}, we always stick to this conditional probability path as the default setting throughout the paper.

\subsection{Problem Formulation}\label{sec:problem_formulation}
We now describe the music generation task and the general methodology based on flow matching that we employ.
Given a dataset consisting of audio-text pairs $(x, w)$, where $x\in\mathbb{R}^{T\times C}$ ($T$: number of timesteps, $C$: number of channels) is the music audio and $w=\{w_1,w_2,..,w_n\}$ ($w$: words) 
is the corresponding textual description represented as a sequence of words, the goal is to build a text-conditioned music generation model $p(x|w)$. In addition to generating music from scratch, we further consider two practical tasks: music continuation 
$p(x_{t_1:T}|x_{1:t_1}, w)$
and music infilling $p(x_{t_1:t_2}|x_{1:t_1}, w, x_{t_2:T})$, with $t_1$, $t_2\in[0,T]$.
In order to allow the model to perform all the text-guided music generation, we formulate our approach as an in-context learning task following~\cite{Le2023VoiceboxTM}. Specifically, given a binary temporal mask $m$ for a music track $x$, we train a conditional flow matching model predicting the vector field in the masked regions of the music track $x_{m} = x \odot m$ while conditioning on the
 unmasked regions of the music track $x_{ctx} = x \odot (1 - m)$ and the text caption $w$ about the music piece. Formally, we train with the following flow matching loss: $L_{\text{CFM}}=\mathbb{E}_{t,m,p(x|x_1),q(x_1,w)}||m\odot(v_t(x,x_{ctx},w;\theta)-u_t(x|x_{ctx},w))||^2$.
 
 In addition to increasing the model capacity, such masked prediction objective also benefits generative modeling in general, as is shown in~\cite{jen1,Le2023VoiceboxTM}. Within this framework, the three tasks of TTM, music continuation and music infilling can be conceptualized as setting specific mask values for $p(x_{m}|(1-m)\odot x,w)$, where $m$ is set to be $\mathbf{1}_{1:T}$, $[\mathbf{0}_{1:t_1},\mathbf{1}_{t_1:T}]$ and $[\mathbf{0}_{1:t_1},\mathbf{1}_{t_1:t_2},\mathbf{0}_{t_2:T}]$ respectively.

\subsection{A Cascaded Flow-matching Approach}
\begin{figure*}
    \centering
    \includegraphics[width=0.75\linewidth]{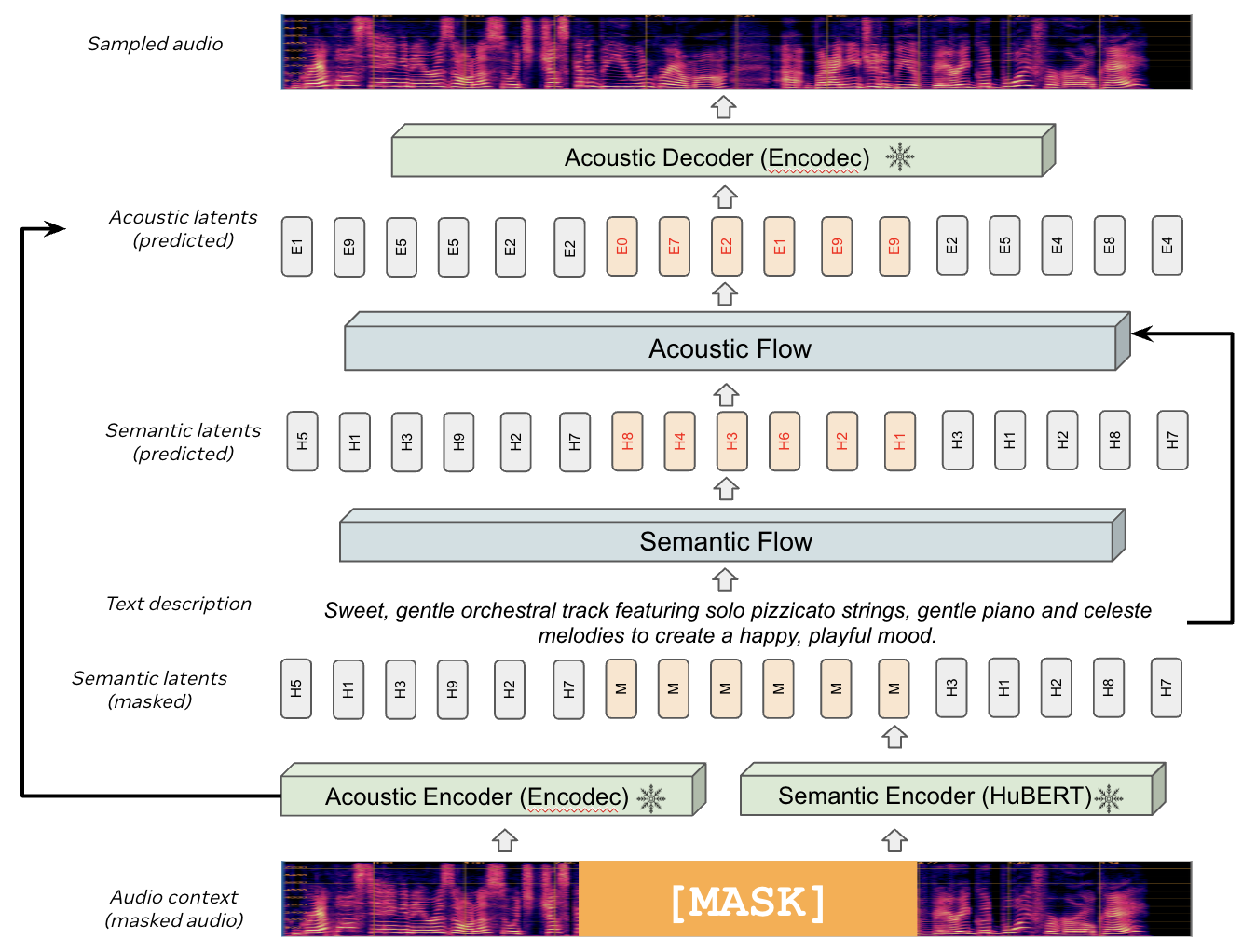}
    \caption{MusicFlow Diagram. Note the acoustic encoder, acoustic decoder and semantic encoder are pre-trained and frozen during generative model training. For text-to-music generation (i.e., 100\% masking), both acoustic and semantic encoder are discarded in inference.
    }
    \label{fig:main_fig}
\end{figure*}
Training flow matching to directly generate music conditioned on text captions is difficult~\cite{audio-ldm2} given the vast number of potential music tracks corresponding to a single text caption. As the text caption lacks the fine-grained information to adequately describe a music track, we propose to condition on a latent music representation that describes the music at the frame level.

MusicFlow is thus divided into two stages: semantic modeling and acoustic modeling. The first stage outputs latent representations $h = (h_1, h_2,....h_M) \in \mathcal{R}^{M\times D}$ conditioned on a text caption $w$. In the second stage, we condition on the latent representations from the first-stage model, and a text caption $w$ to output low-level acoustic features of $N$ frames, $x = (x_1, x_2, ...., x_N) \in \mathcal{R}^{N\times C}$. Note $h$ and $x$ are monotonically aligned. Both stages are inherently stochastic, meaning there are multiple potential $(h,x)$ pairs for a given caption $w$. Therefore, we model these two stages separately with flow matching. In both stages, we predict masked vector fields as discussed before and provide detailed descriptions on the two stages below.

\subsection{Stage 1: Music Semantic Flow Matching from Text}

Our first-stage model consists of generating the semantics of the music piece conditioned on the text description. Here the semantics refer to high-level musical information instead of fine-grained details such as the general audio quality, which are inferred from the text description. For music, the semantics can refer to the melody and rhythm or harmony in a piano piece, analogous to the linguistic content in speech. 

\textbf{Semantic latent representation}
One natural way of representing music is through music transcription.
Transcripts in music typically refer to some notation system (e.g., music scores) that indicates the pitches, rhythms, or chords of a musical piece.
A notable advantage of music transcript is its interpretability as it is human-readable and thus poses easy alignment with humans. However, for large-scale audio datasets, the associated music transcripts are usually not readily available, while manual annotation involves a non-trivial amount of labeling efforts. Automatic music transcription is a challenging task~\cite{Benetos2019Automatic} and the existing approaches~\cite{Bittner2022ALI,Hawthorne2021SequencetoSequencePT,hsu2021vocano,Su2019tent,Hawthorne2018EnablingFP} are heavily restricted to a single-instrument setting (e.g.
piano, solo vocals, etc.). 

To  address the challenge of acquiring music transcriptions, we adopt HuBERT~\cite{Hsu2021HuBERTSS}, a popular self-supervised speech representation learning framework, to obtain frame-level semantic features, which can be regarded as a form of pseudo transcription. In essence, a HuBERT model consists of masked prediction of hidden units from the raw audio, which are inferred initially from MFCC and iteratively refined with layerwise features. For speech, HuBERT units have shown to correlate well with phonemes~\cite{Hsu2021HuBERTSS} and its intermediate features entails rich semantic information~\cite{Pasad2023comparative}. In music understanding tasks, HuBERT has been successfully applied in source separation~\cite{Pasini2023SelfSupervisedMS}, shedding light on its potential for capturing musical characteristics. As the original HuBERT model is pre-trained with speech only, we re-train HuBERT using music data following the original recipe. Training details are given in Section~\ref{sec:exp}.

\textbf{Semantic flow matching} Given a HuBERT model $\mathcal{H}$, one can extract the semantic features from its $l$th layer $l$: $h = \mathcal{H}(x) \in \mathcal{R}^{M\times C_h}$, where $C_h$ is the HuBERT feature dimension. The layer index $l$ is tuned in practice. A text-conditioned semantic flow-matching model $p(h|w)$ can be trained given text-feature pairs $(h, w)$. As described in Section~\ref{sec:problem_formulation}, we adopt the masked prediction objective by conditioning on the context $h_{ctx}=m\odot h$, where $m$ is a span mask of length $M$. More formally, we adopt the following training objective for the semantic modeling stage: $L_{\text{H-CFM}}=\mathbb{E}_{t,m,p(h|h_1),q(h_1,w)}||m\odot(v_t(h,h_{ctx},w;\theta)-u_t(h|h_{ctx},w))||^2$.
Cross-attention layers are integrated into the backbone model, enabling it to attend to the text description $w$, akin to~\cite{rombach2021highresolution}. 

As an alternative to modeling the distribution of dense features, one can quantize the layerwise features $h$ into units $u$ and model the unit distribution $p(u|w)$ instead. For this case, a straightforward method is to build an autoregressive language model, which factorizes $p(u|w)=\displaystyle\prod_{n=1}^{M}p(u_{n}|u_{1:{n-1}},w)$. Using a semantic LM has been explored in~\cite{musiclm} in hierarchical LMs for music generation. We also noticed its effectiveness when combined with flow matching, as will be shown in Section~\ref{sec:exp}. However, this hybrid model is unsuitable for music infilling task due to its left-to-right nature.

\subsection{Stage 2: Music Acoustic Flow Matching from text and semantics}
\textbf{Acoustic latent representation} The second-stage model aims to infer the low-level acoustic information (e.g., volume, recording quality) implied by the semantic tokens. Directly predicting raw waveforms ensures the completeness of information while imposes the challenge of modeling the long sequences. To balance between quality and sequence length, we use Encodec~\cite{defossez2022highfi} to map raw waveforms into dense feature sequences. In a nutshell, Encodec~\cite{defossez2022highfi}, an auto-encoder based on residual vector quantization, comprises of an encoder $\mathcal{E}$ and decoder $\mathcal{D}$. During training, we map raw waveforms into acoustic features with the encoder $\mathcal{E}$: $e=\mathcal{E}(x)\in\mathbb{R}^{N\times C_e}$, where $C_e$ is the feature dimension of encodec.

\textbf{Acoustic flow matching} The second-stage flow matching aims to model the following conditional distribution: $p(e|h, w)$. Similar to semantic flow matching, we apply masked prediction and the corresponding training objective is formulated as: $L_{\text{E-CFM}}=\mathbb{E}_{t,m,p(e|h_1, e_1),q(e_1, h_1,w)}||m\odot(v_t(e,e_{ctx},h_1,w;\theta)-u_t(e|e_1,h_1,e_{ctx},w))||^2$. 
As the semantic and acoustic features are aligned ($N/M\approx{sr}_\mathcal{E}/{sr}_\mathcal{H}$, $sr$: sample rate), we simply linearly interpolate the HuBERT feature sequence $h$ to length $N$ before feeding it into encoding. 

Note though Encodec includes multiple different codebooks to quantize the latent features, we directly model the dense feature sequence from the encoder $\mathcal{E}$ without any quantization. This avoids the length increase brought by using multiple codebooks, where the total number of discrete tokens is $K-1$ times more than the dense feature length. Thus, it eliminates the necessity of carefully designing interleaving pattern of discrete tokens to account for dependencies between multiple codebooks~\cite{musicgen,Wang2023NeuralCL}.


\subsection{Classifier-free guidance}

During inference, we sequentially sample the HuBERT features $\hat{h}$ and encodec features $\hat{e}$ using the estimated vector field $v_t(h,h_{ctx},w;\theta_h)$ and $v_t(e,e_{ctx},\hat{h},w;\theta_e)$ following the ODE equation~\ref{eq:ode}. 
The acoustic features are decoded into waveforms via the decoder $\mathcal{D}$ of the Encodec.

As is common in diffusion models, classifier-free guidance is a widely used technique to balance sample diversity and text coherence. Thus we also adopt it in our cascaded generation framework. For flow matching, using classifier-free guidance~\cite{Zheng2023GuidedFF} consists of computing a linear combination between conditional and unconditional vector field: $\tilde{v}^{H}_t(h, w, h^{ctx};\theta_{h}) = (1 + \alpha_h)v^{H}_t(h, w, h^{ctx};\theta_h) - \alpha_h v^{H,uncond}_t(h;\theta_h)$ and $\tilde{v}^{E}_t(e,e^{ctx},\hat{h},w;\theta_{e}) = (1 + \alpha_e)v^{e}_t(e,e^{ctx},\hat{h},w;\theta_{e}) - \alpha_e v^{e,uncond}_t(e;\theta_e)$. 

In order to model the unconditional vector field $v_t^{H,uncond}$ and $v_t^{E,uncond}$ with $v_t^H$ and $v_t^E$ , we randomly drop the conditions (e.g., text, the contextual features) in both flow models with probability $p^{H}$ and $p^{E}$ in training, whose values are also tuned.

%

%



\section{Experiments}
\label{sec:exp}

\begin{table*}
\caption{Comparisons between MusicFlow with previous works in text-to-music generation on the MusicCaps dataset. }
\label{tab:main-result}
\begin{center}
\begin{sc}
\resizebox{\textwidth}{!}{
\begin{tabular}{lcccccc}
\toprule
Model & \# params & FAD($\downarrow$) & FD($\downarrow$) & KL-div($\downarrow$) & ISc.($\uparrow$) &  CLAP-text($\uparrow$)\\
\midrule
MusicLM~\cite{musiclm} & 860M & 4.00 & - & - & -  & - \\
MusicGen~\cite{musicgen} & 1.5B & 3.40 & 24.1 & 1.23 & 2.29  & 0.37\\
UniAudio~\cite{yang2023uniaudio} & 1B & 3.65 & - & 1.90 & - & -\\
AudioLDM-2~\cite{audio-ldm2} & 746M & 3.13 & 18.8 & \textbf{1.20} & 2.77 & 0.43\\
Noise2Music~\cite{noise2music} & 1.3B & 2.10 & - & - & - & -\\
JEN-1~\cite{jen1} & 746M     & \textbf{2.00} & - & 1.29 & -  & -       \\
\midrule
MusicFlow (unidirectional LM + FM)  & 546M & 2.69 & \textbf{13.2} & 1.23 & 2.69 & 0.52 \\
MusicFlow (bidirectional  FM + FM) & 330M & 2.82 & 14.2 & 1.23 & \textbf{2.78} & \textbf{0.56} \\
\bottomrule
\end{tabular}
}
\end{sc}
\end{center}
\end{table*}

%
%
%

%

\subsection{Experimental Setup}
\textbf{Data}
We use 20K hours of proprietary music data ($\sim$400K tracks) to train our model. We follow the original recipe in~\cite{Hsu2021HuBERTSS} to train the music HuBERT model with music data. For data preprocessing, we filter out all the vocal tracks and resample all the data to 32kHz and perform channel-wise averaging to downmix all multi-channel music into mono. Only text descriptions are retained for training, while the other metadata such as genre, BPM and music tags are discarded.
We evaluate our model on MusicCaps~\cite{musiclm}, which incorporates 5.5K
10s-long audio samples annotated by expert musicians in total. For subjective evaluation, we use the 1K genre-balanced subset following~\cite{musiclm}.

\textbf{Implementation details} We follow~\cite{Le2023VoiceboxTM} for backbone architectures in both stages, which are Transformers~\cite{Vaswani2017AttentionIA} with convolutional position
embeddings~\cite{Baevski2020wav2vec2A}, symmetric bi-directional ALiBi self-attention bias~\cite{Press2021TrainST} and UNet-style skip connections. Specifically, the transformers of the first and second stage include 8 and 24 layers of 12 attention heads with 768/3072 embedding/feed-forward network (FFN) dimension, leading to 84M and 246M
parameters (see Section~\ref{sec:ablation-model-size} for ablation on model size). 
The models are trained with an effective batch size of 480K frames, for 300K/600K updates in two stages respectively. For efficiency, audios
are randomly chunked to 10s during training. For masking, we adopt the span masking strategy and the masking ratio is randomly chosen between $70-100\%$. Condition dropping probabilities (i.e., $p^H$ and $p^E$) are 0.3 for both stages. 
We use the Adam optimizer~\cite{kingma2014adam} with learning rate 2e-4, linearly warmed up for 4k steps
and decayed over the rest of training. 

\textbf{Objective evaluation} We evaluate the model using the standard Frechet Audio Distance (FAD)~\citep{Kilgour2019FrchetAD}, Frechet Distance (FD) and KL divergence (KLD) based on the pre-trained audio event tagger PANN~\citep{Kong2019PANNsLP}, and Inception score (ISc)~\citep{Salimans2016ImprovedTF}, which are adapted from sound generation and has been widely used in prior works for text-to-music generation~\cite{musiclm,noise2music,musicgen,jen1}.
Specifically, FAD and FD measure distribution-level similarity between reference samples and generated samples. KLD is an instance level metric computing the divergence of the acoustic event posterior between the reference and the generated sample for a given description. 
The metrics are calculated using the \texttt{audioldm\_eval} toolkit.\footnote{\url{https://github.com/haoheliu/audioldm_eval}}. 
To measure how well the generated music matches the text description, we use CLAP~\footnote{We use the \url{music_speech_epoch_15_esc_89.25.pt} checkpoint, trained on both speech and music data.} similarity, defined as the cosine similarity between audio and text embeddings.

\bsc{This paragraph is mostly copied from AudioBox, and needs some rewording.}

\textbf{Subjective evaluation} In addition to the objective metrics mentioned above, we further conduct subjective evaluation with human annotators. The study consists of multiple pairwise studies following the evaluation protocol of~\citet{musiclm}. Specifically, each human annotator is presented with pairs of audio clips generated by two different systems and is required to give their preference based on how well the generated music captures the elements in the text description.

\subsection{Main Results}\label{sec:main-result}

Table~\ref{tab:main-result} compares our model to prior works in text-to-music generation on MusicCaps in terms of objective metrics. 
Given the variation in models used for evaluation in prior works, we primarily rely on FAD, which is computed using the vggish feature~\cite{Kilgour2019FrchetAD} and serves as a unified benchmark across different studies. Specifically, when evaluating MusicGen, we opt for its medium version due to its overall superior performance compared to other variants~\cite{musicgen}. For MusicGen and AudioLDM2, we use the public model checkpoints in order to get FD, ISc and CLAP similarity since these metrics were not reported in the paper. 

In MusicFlow, we additionally present the results of a model with the first stage functioning as a language model, predicting HuBERT units as detailed in Section~\ref{sec:method}. The language model includes 24 transformer layers, 16 attention heads, and a hidden dimension of 1024, leading to $\sim300M$ parameters in total.

\begin{table*}
\label{tab:infil-cont}
\caption{Performance of MusicFlow on various music generation tasks on MusicCaps dataset. We compare with AudioLDM-2~\cite{audio-ldm2} for text-to-music and AudioLDM for music infilling and continuation.}
\vspace{-0.2in}
\begin{center}
\begin{sc}
\resizebox{\textwidth}{!}{%
\begin{tabular}{lccccccc}
\toprule
Task // Model  & FAD($\downarrow$) & FD($\downarrow$) & KL-div($\downarrow$) & ISc.($\uparrow$) & CLAP-sim($\uparrow$) & CLAP-audio($\uparrow$) & CLAP-text($\uparrow$) \\
\midrule

\rowcolor{verylightgray}
\multicolumn{8}{c}{\textbf{Text-to-music (100\%)}} \\
AudioLDM-2~\cite{audio-ldm2} & 3.13 & 18.8 & \textbf{1.20} & 2.77 & - & 0.44 & 0.43 \\
MusicFlow & \textbf{2.82} & \textbf{14.2} & 1.23 & \textbf{2.78} & - & \textbf{0.48} & \textbf{0.56} \\
\midrule
\rowcolor{verylightgray}
\multicolumn{8}{c}{\textbf{Continuation (last 70\%)}} \\
AudioLDM~\cite{liu2023audioldm} & 2.08 & 25.08 & 0.66 & 2.80 & 0.61 & 0.61 & 0.53 \\
MusicFlow & \textbf{1.63} & \textbf{6.50} & \textbf{0.49} & \textbf{3.37} & \textbf{0.88} & \textbf{0.77} & \textbf{0.56} \\
\midrule
\rowcolor{verylightgray}
\multicolumn{8}{c}{\textbf{Infilling (middle 70\%)}} \\
AudioLDM~\cite{liu2023audioldm} & 2.09 & 45.93 & 0.76 & 2.39 & 0.59 & 0.61 & 0.54 \\
MusicFlow & \textbf{1.71} & \textbf{6.5} & \textbf{0.38} & \textbf{3.18 }& \textbf{0.89} & \textbf{0.79} & \textbf{0.57} \\
\bottomrule
\end{tabular}
}
\end{sc}
\end{center}
\end{table*}

In comparison to all prior works, our model exhibits a significant reduction in size, with parameter reduction ranging from 50\% to 80\%, while remaining competitive in terms of generation quality. Compared with a standard diffusion model - AudioLDM-2, MusicFlow achieves a 10\% lower FAD ($3.13\rightarrow2.82$) with approximately 50\% fewer parameters. Similarly, compared to the language-model-based MusicGen, our approach shows a 20\% improvement in FAD ($3.40\rightarrow2.82$) while using only 20\% of the parameters. These results highlight the efficiency of our approach.

It's noteworthy that MusicLM~\cite{musiclm} shares similarities with ours, incorporating semantic and acoustic modeling stages based on language models. However, we surpass this approach by roughly 30\% in FAD with less than 65\% of the parameters. Additionally, in contrast to the current state-of-the-art model on MusicCaps, Jen-1~\cite{jen1}, our results shows a mixture of results. While falling behind in FAD, we outperform it in KL divergence with only half of the parameters.

\textbf{LM vs. FM for first stage} In addition to our main approach, we investigate the integration of a language model for first-stage modeling. Both approaches share the second-stage model. According to the last two rows of table~\ref{tab:main-result}, using a first-stage LM yields marginally superior results compared to using a flow matching model. This implies that semantic features in music audios possess discrete structures, which can be well captured by an auto-regressive language model. Nonetheless, for the sake of model efficiency and task generalization, we adhere to using the flow matching cascade moving forward.   

\begin{figure*}
    \centering
    \label{fig:human-eval}
    \includegraphics[width=0.9\linewidth]{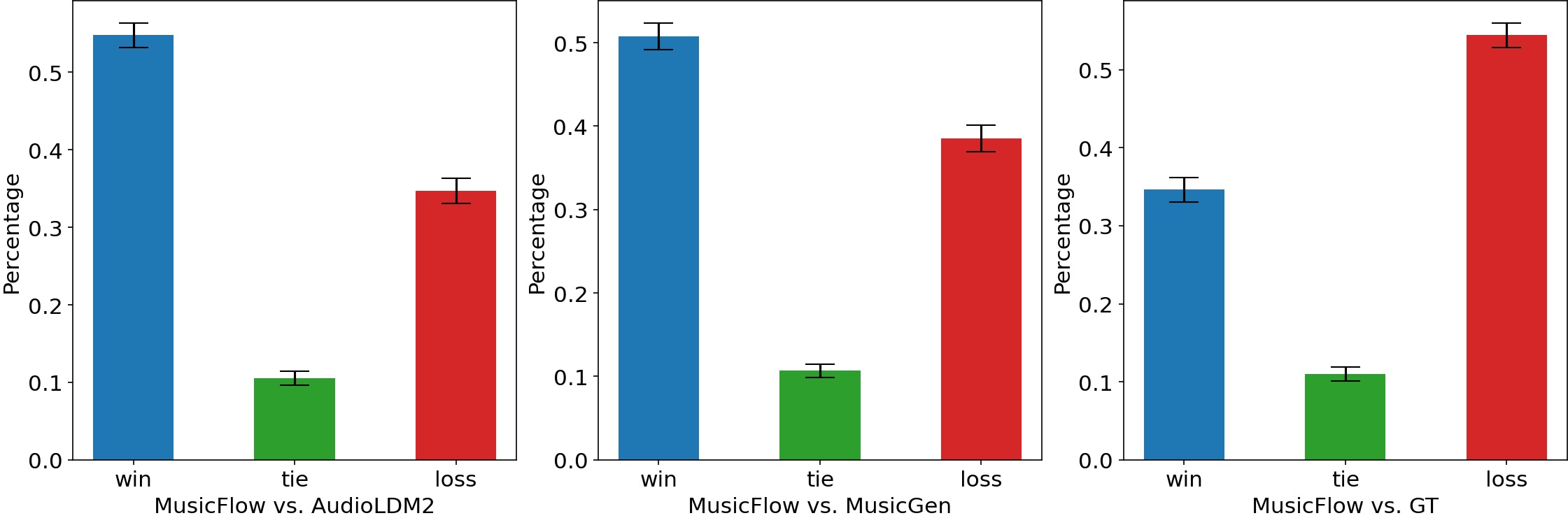}
    \vspace{-0.1in}
    \caption{Pairwise comparison between MusicFlow, AudioLDM2, MusicGen and ground-truth}
\end{figure*}

\textbf{Subjective evaluation} Figure~\ref{fig:human-eval} shows the pairwise comparison between our model and prior works. In particular, we compare MusicFlow to AudioLDM2 and MusicGen, which are the only two publicly available models in Table~\ref{tab:main-result}.
For our model, we use the \texttt{bidirectional FM+FM} configuration in Table~\ref{tab:main-result}.
Our model surpasses both AudioLDM2 and MusicGen.
This observation aligns with the objective metrics presented in Table~\ref{tab:main-result}. However, it's worth noting that there is still a gap between our model and the ground-truth.

\begin{figure}[htp]
    \centering
    \includegraphics[width=\linewidth]{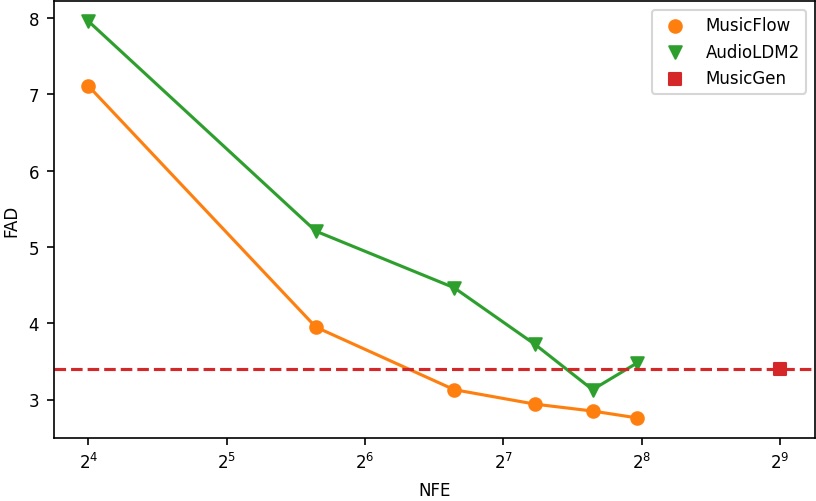}
    \vspace{-0.3in}
    \caption{Comparison between MusicFlow and prior works in FAD-NFE in terms of inference efficiency.}
    \label{fig:fad-nfe}
\end{figure}

\textbf{Inference Efficiency} In Table~\ref{tab:main-result}, we only lists the model size, which is one aspect of model efficiency.  In Figure~\ref{fig:fad-nfe}, we plot how FAD changes when we vary the number of function evaluations (NFE) during inference. For flow matching and AudioLDM2, this is achieved by adjusting the number of iterative steps in the ODE solver\footnote{We use the \texttt{midpoint} solver for this analysis} and DDIM steps, respectively. Since MusicFlow involves two flow matching models, we simply aggregate the NFE of the two modules as the final NFE we plot. For comparison, we further show the MusicGen, which runs a fixed number of auto-regressive steps. As shown in Figure~\ref{fig:fad-nfe}, MusicFlow outperforms MusicGen (FAD: 3.13 vs. 3.40) by using $20\%$ of inference steps. Running with longer steps further improves the performance. The final model takes only $50\%$ the network forward passes of MusicGen. 
AudioLDM2 exhibits a similar trend to ours, although its generation quality consistently lags behind with the same number of inference steps.

\subsection{Infilling and Continuation}
One advantage of MusicFlow is its ability to handle multiple audio-conditioned generative tasks, such as infilling and continuation, with a single model. These tasks have also been explored in~\cite{jen1}, albeit without reported quantitative metrics.  Due to lack of baselines, we compare the model performance to the our own text-to-music model, as detailed in Table~\ref{tab:main-result}. For the infilling task, we infill the middle 70\% of the audio segment. For the continuation task, given the beginning 30\% of the audio clip, the model generates the remaining 70\%.  

As is shown in Table~\ref{tab:infil-cont}, our model effectively uses the context to enhance audio generation. In both settings, using a 3s audio context enables a nearly 50\% reduction in FAD. The text-to-audio similarity is slightly increased in infilling ($0.44\rightarrow0.45$). 
We hypothesize this may be because the CLAP model struggles to discern fine-grained details in the text description. Hence, we conduct a subjective study to measure text faithfulness. 
The MOS scores of text-to-music, conditnuation and infilling are respectively $3.34\pm 0.18$, $3.47\pm 0.18$, $3.42\pm 0.19$ with 95\% confidence interval. This confirms an improvement in text faithfulness through context utilization.

Additionally, among other metrics, we compute the CLAP-Audio score, defined as the cosine similarity between the embeddings of the generated and ground-truth audios. Compared to text-only generation, the generated audio achieved higher scores, suggesting better acoustic matching through context conditioning.  
Finally, we measure the CLAP similarity between the generated segment and the original context (CLAP-SIM). Both settings achieve scores close to 1, implying coherence between the generation and context.

\subsection{Ablation Study}
Below we analyze the impact of different design choices in MusicFlow, particularly focusing on the necessity of a two-stage cascade and how model scales differently in each stage.

\subsubsection{Single-stage vs. Two-stage Model} 
We compare MusicFlow to a simple flow matching baseline of directly generating music based on text descriptions without the intermediate HuBERT features in Table~\ref{tab:2vs1}. Including a stage of HuBERT prediction consistently improves the performance across various metrics regardless of model size. HuBERT-based flow matching brings a $\sim 30\%$ relative improvement in terms of FAD. 
Note while we increased the size of the single-stage model to 431M, it did not yield additional gains, despite having more parameters.


%

\begin{table}[h]
\caption{Comparison between single-stage and multi-stage flow matching in different model sizes}
\label{tab:2vs1}
\begin{center}
\begin{small}
\begin{sc}
\begin{tabular}{lcccc}
\toprule
Model & FAD & FD & KL-div & ISc.  \\
\midrule
Single-stage (123M) & 4.58 & 25.5 & 1.62  & 2.52 \\
Single-stage (246M) & 4.52 & 22.9 & 1.57  & 2.66 \\
Single-stage (431M) & 5.11 & 27.5 &  1.64 & 2.68 \\
Two-stage (84M+123M) & 3.37 & 20.6 & 1.50  & 2.59 \\
Two-stage (84M+246M) & \textbf{2.82} & \textbf{14.2} & \textbf{1.23}& \textbf{2.78}\\
\bottomrule
\end{tabular}
\end{sc}
\end{small}
\end{center}
\end{table}

\subsubsection{Effect of model size}
\label{sec:ablation-model-size}
Empirically, we observed the performance of our models is heavily influenced by the model size. In this analysis, we delve into the impact of model size in each stage.

\noindent\textbf{Second-stage: Text + HuBERT to Music.} We first examine how the size of the second-stage model, specifically the Text + HuBERT features $\rightarrow$ music, affects the overall performance. We keep the best first-stage model and scale the second-stage model by altering the number of transformer layers and the hidden dimension of each layer (see Table~\ref{tab:second-scaling}). The performance improves as we increase the number of layers until reaching 24 layers. 
Beyond this point, increasing the number of layers or feature dimensions results in degradation, suggesting a potential overfitting issue of the model.

\noindent\textbf{First-stage: Text to HuBERT.} We fix the configuration for our second-stage model based on the above findings and vary only the first-stage configuration (see Table~\ref{tab:first-scaling}).
Unlike the second-stage, where the best model is with 24 transformer layers, our best first-stage model for Text $\rightarrow$ HuBERT feature prediction is notably smaller with an optimal configuration of only 8 layers. According to Table~\ref{tab:first-scaling}, smaller models typically perform equally well or even better than their larger counterparts in the first-stage model. We hypothesize that predicting HuBERT features is simpler than predicting the low-level Encodec features, particularly for shorter music pieces with standard music structures, as the former consists of learning only the coarse-grained semantics. 
Consequently, a larger variant is more susceptible to overfitting compared to the second-stage scenario.

\begin{table}[htp]
\caption{Effect of the \emph{second-stage} model size  on performance. In each row we specify the number of layers, the hidden dimension of the transformer and the total number of trainable parameters.}
\label{tab:second-scaling}
\begin{center}
\begin{small}
\begin{sc}
\begin{tabular}{lcccc}
\toprule
Model configuration & FAD & FD & KL-div & ISc. \\
\midrule
12L, 768d (123M) & 3.37 & 20.6 & 1.50  & 2.59 \\
18L, 768d (123M) & 3.22 & 18.2  & 1.42  & 2.62 \\
24L, 768d (246M) & \textbf{2.82} & \textbf{14.2} & \textbf{1.23}& \textbf{2.78} \\
32L, 768d, (323M) & 3.12 & 17.9 & 1.42 & 2.64  \\
\midrule
12L, 1024d, (217M) & 3.56 & 18.7 & 1.43 & 2.67\\
18L, 1024d, (324M) & 3.26 & 18.4 & 1.42  & 2.67  \\
24L, 1024d, (441M) & 3.40 & 17.8 & 1.43 & 2.71 \\
\bottomrule
\end{tabular}
\end{sc}
\end{small}
\end{center}
\end{table}

\begin{table}[htp]
\caption{Effect of the \emph{first-stage} model size  on performance. In each row we specify the number of layers, the hidden dimension of the transformer and the total number of trainable parameters.}
\label{tab:first-scaling}
\begin{center}
\begin{small}
\begin{sc}
\begin{tabular}{lcccc}
\toprule
Model configuration & FAD & FD & KL-div & ISc. \\
\midrule
12L, 1024d (217M) & 3.18 & 18.2 & 1.44  & 2.74 \\
12L, 768d (123M) & 3.09 & 17.1 & 1.42 & 2.73 \\
8L, 768d (84M) & \textbf{2.82} & \textbf{14.2} & \textbf{1.23}& \textbf{2.78} \\
6L, 768d, (64M) & 3.30 & 18.1 & 1.47  & 2.69  \\
8L, 512d, (38M) & 3.20 & 17.6 & 1.41 & 2.76 \\
\bottomrule
\end{tabular}
\end{sc}
\end{small}
\end{center}
\end{table}

\subsubsection{Number of training iterations} 
We notice the performance of both stages in MusicFlow is sensitive to the number of training iterations. Generally, longer training boosts performance, as can be seen from Table~\ref{tab:iter-tuning}. While varying the number of training iterations, we maintains the sizes of best models from Table~\ref{tab:first-scaling} and~\ref{tab:second-scaling}. Comparing the two stages, longer training consistently enhances performance in the second stage, while there is a degradation in performance with further increases in training iterations in the first stage. 
This aligns with our observations in model scaling, which highlight the different tendencies of model overfitting in both stages.

\begin{table}[h]
\caption{Impact of training steps on the model performance}
\label{tab:iter-tuning}
\begin{center}
\begin{small}
\begin{sc}
\begin{tabular}{cccccc}
\toprule
Stage 1 & Stage 2 & FAD & FD & KL-div & ISc. \\
\midrule
100K & \multirow{4}{*}{600K} & 3.60 & 18.1 & 1.42  & 2.54 \\
200K & & 3.00 & 17.7  & 1.45 & 2.79 \\
300K & & \textbf{2.82} & \textbf{14.2} & \textbf{1.23}& \textbf{2.78} \\
400K & & 2.90 & {16.3} & {1.39} & 2.85  \\
\midrule
\multirow{3}{*}{300K} & 200K & 3.19 & 19.3 & 1.42 & 2.51 \\
 & 400K & 2.84 & 16.6 & \textit{1.39} & 2.71 \\
  & 600K & \textbf{2.82} & \textbf{14.2} & \textbf{1.23}& \textbf{2.78} \\
\bottomrule
\end{tabular}
\end{sc}
\end{small}
\end{center}
\vskip -0.1in
\end{table}

\subsubsection{Choice of Semantic Latent Representation} 
The first stage model predicts semantic latent representations conditioned on text tokens. The choice of the semantic latents has an impact on  the final performance. In addition to HuBERT units, we also experiment with MERT units~\cite{mert} using the officially released pre-trained music model. In Table~\ref{tab:semantic_latents}, we can see that it is clearly worse compared to using HuBERT units.

\begin{table}[htp]
\caption{Impact of using a different semantic latent representation instead of HuBERT. We compare with MERT~\cite{mert} units below.}
\label{tab:semantic_latents}
\begin{center}
\begin{small}
\begin{sc}
\begin{tabular}{lcccc}
\toprule
Semantic Latent & FAD & FD & KL-div & ISc. \\
\midrule
MERT & 3.43 & 18.3 & 1.47  & 2.54 \\
HuBERT & \textbf{2.82} & \textbf{14.2} & \textbf{1.23}& \textbf{2.78} \\
\bottomrule
\end{tabular}
\end{sc}
\end{small}
\end{center}
\vskip -0.1in
\end{table}

\subsubsection{Choice of Acoustic Latent Representation} 
The second stage model predicts acoustic latent representations from the semantic latent features. The choice of the acoustic latents also affects the final performance. In addition to Encodec, we also experiment with the recently proposed UniAudio tokenizer~\cite{uniaudio} and DAC~\cite{dac}. We could not achieve convergence with DAC, and found UniAudio to perform slightly worse compared to Encodec in terms of all quantitative metrics. We report the results in Table~\ref{tab:acoustic_latents}.

\begin{table}[htp]
\caption{Impact of using a different acoustic latent representation instead of Encodec. We compare with UniAudio~\cite{uniaudio} below. We could not achieve convergence with DAC~\cite{dac}}
\label{tab:acoustic_latents}
\begin{center}
\begin{small}
\begin{sc}
\begin{tabular}{lcccc}
\toprule
Acoustic Latent & FAD & FD & KL-div & ISc. \\
\midrule
DAC & - & - & - & - \\
UniAudio & 3.18 & 18.2 & 1.44  & 2.74 \\
Encodec & \textbf{2.82} & \textbf{14.2} & \textbf{1.23}& \textbf{2.78} \\
\bottomrule
\end{tabular}
\end{sc}
\end{small}
\end{center}
\end{table}

\section{Conclusion}
We present MusicFlow, a cascaded flow-matching network for text-guided music generation. Our model leverages a self-supervised model to capture semantic information within music audio. 
Comprising two flow matching networks that predict semantic and acoustic features in a cascaded manner, MusicFlow consistently outperforms all public text-to-music models in both subjective and objective metrics, with only a fraction of model parameters and inference steps.
Overall, MusicFlow achieves performance on par with the state-of-the-art models while being significantly smaller. 
Additionally, our model allows text-guided music continuation and infilling through in-context learning, eliminating the need for task-specific training. Our future work includes further improving model efficiency by using sophisticated ODE solvers such as~\cite{shaul2023bespoke}. 


\section*{Impact Statement}
While music generation technologies make music creation more accessible to amateur creators, they also pose potential societal challenges. Given that modern music generation models often require substantial data, preventing copyright infringement deserves careful attention. In this work, we ensure the use of music data for model training adheres to legal terms. For future data scaling, it's essential to inform artists of data usage and provide opt-out options, as commonly practiced in concurrent music generation works. Furthermore, we acknowledge the lack of diversity in our model generations, potentially stemming from the predominantly stock music training data with limited world music. Our future objective is to ensure high-quality music generation across diverse genres.

\bibliography{main}
\bibliographystyle{icml2023}

\end{document}